\documentclass[aip]{revtex4-1}
\usepackage{graphicx, amsmath, amssymb}

\begin{document}

\title{Sub-THz complex dielectric constants of smectite clay thin samples with Na$^{+}$/Ca$^{++}$-ions}

\author{Rezwanur Rahman}
\affiliation{Department of Physics, Colorado School of Mines, Golden, CO 80401-1887, USA}
\affiliation{Department of Petroleum Engineering, Colorado School of Mines, Golden, CO 80401-1887, USA}
\affiliation{OCLASSH, Department of Petroleum Engineering, Colorado School of Mines, Golden, CO 80401-1887, USA}
\author{Douglas K. McCarty}
\affiliation{Chevron ETC, 3901 Briarpark, Houston, TX 77402, USA}
\author{Manika Prasad}
\affiliation{Department of Petroleum Engineering, Colorado School of Mines, Golden, CO 80401-1887, USA}
\affiliation{OCLASSH, Department of Petroleum Engineering, Colorado School of Mines, Golden, CO 80401-1887, USA}

\begin{abstract}

    We implement a technique to
    characterize the electromagnetic properties at frequencies 100 to 165
    GHz (3 cm$^{-1}$ to 4.95 cm$^{-1}$) of oriented smectite
    samples  using an open cavity resonator connected
    to a sub-millimeter wave VNA (Vector Network Analyzer).
    We measured dielectric constants perpendicular to the
    bedding plane on oriented Na$^{+}$ and Ca$^{++}$-ion stabilized
    smectite samples deposited on a glass slide at ambient
    laboratory conditions (room temperature and room light).  The clay
    layer is much thinner ($\sim$ 30 $\mu$m) than the glass
    substrate ($\sim$ 2.18 mm).  The real part of dielectric constant,
    $\epsilon_{re}$, is essentially constant over this frequency
    range but is larger in
    Na$^{+}$- than in Ca$^{++}$-ion infused clay.  The total electrical conductivity
    (associated with the imaginary part of dielectric constant,
    $\epsilon_{im}$) of both samples increases monotonically at lower
    frequencies ($<$ 110 GHz), but shows rapid increase
    for Na$^{+}$ ions in the regime $>$ 110 GHz.
    The dispersion of the samples
    display a dependence on the ionic strength in the clay
    interlayers, i.e., $\zeta$-potential in the Stern layers.

\end{abstract}

\maketitle

\section{Introduction}
Clay minerals have a complex layered structure with exchangeable
cations that can bind water molecules in the interlayers. With
increasing pressure and temperature, or in the presence of polar free
radicals, these interlayer cations can be exchanged. This cation
excahange capacity (CEC) of clay minerals affects their
fluid conductivity,\citep{RT86} and permeability\citep{RT86};
dielectric permittivity.\citep{WC82} Complex dielectric properties of clay are
crucial to determine hydrocarbon-contents in oil-rich rocks. These
measurements have been made predominantly at frequencies in the
kHz-range\citep{MM77, MC95} and between 0.5 MHz and 1.1
GHz.\citep{RT86} Complex conductivity of clayey materials between 1 milli-Hertz
(mHz) and 45 kHz for CEC effects has been characterized and
modeled.\citep{AR13} \citet{BC99} researched dielectric properties of
smectite clay samples in detail,  explaining interlayer
polarization and relaxation mechanisms between 30 kHz-300 MHz. Some
clay minerals can swell due to hydration with water adsorbed in the
interlayer depends on the charge of the interlayer cations. Electrical
measurements can yield cation mobility. Dielectric measurements can be
instrumental to characterize the water absorbed in
smectites\cite{WR68} The conductivities of smectite
clays saturated by monovalent cations has also been
studied in detail.\citep{RC75, FJ65}
These frequency-and temperature-dependent measurements were in the
frequency range between 300 Hz and 10 kHz, and from -150$^{\circ}$C to
+30$^{\circ}$C. In a much higher frequency domain, the 
THz dielectric constants of layered silicates
including muscovite, vermiculite, phlogopite, and biotite, have been
measured by THz-time domain spectroscopy.\citep{MJ09}  Unfortunately this
technique is fairly noisy and low resolution since the THz pulses are very
weak.  So, in this paper we look at Monmorillionite using much higher resolution CW methods
based on harmonic multiplication of phase stabilized
microwaves, electronically generated;
these methods provide very low noise/high dynamic range out to about 1.4 THz
at present.

\section{Methods}

At CSM we use three millimeter wave (or sub-THz) modalities to extract
material properties: (1) a quasi-optical
System,\citep{JS0689,JS0688,NG14} to study bulk properties, (2) a near
field scanning system,\citep{MW09} to measure local properties, and
(3) the open hemispherical cavity resonator,\citep{RR13a,Dudorov95}
for samples that are too thin or too low-loss for quasi-optical
techniques.  In this work, using cavity resonance perturbation,
we extract the complex dielectric
constants of clay-thin films in 100-165 GHz and
investigate electrical properties in the presence of
Ca$^{++}$/Na$^{+}$-ions. We study how these cations influence
conductivity of free carriers, and relaxations. We also compare our
data with low frequency measurements.\citep{RT86}

  We use an open hemispherical cavity resonator with VNA (Vector
  Network Analyzer)\citep{RR13a} to measure electrical properties of
  thin sections of clay-samples with Ca$^{++}$/Na$^{+}$-ions
  infused. The cavity is a structure with two copper mirrors
  positioned at certain distance (the ``cavity length'') without any
  sidewalls. The top mirror is hemispherical and connected to two
  WR-10 waveguide couplers working as a transmitter and a receiver,
  and on the other hand the lower mirror is flat and smaller than the
  upper one in size. We measured the real part of refractive index of
  $\sim$ 1 mm-thick glass substrate (borosillicate) to be 1.98 at 310
  GHz which is the same as its published value.\citep{RR13a} For
  details on the cavity and methodology see \citet{RR13a}.

\subsection{Open Cavity Resonator}

The principle of this technique is cavity perturbation. The changes
in axissymetric mode profiles, mainly the frequency-shift and
linewidth-variation, between an empty cavity mode and the same mode
in presence of a sample, allows us to determine the complex dielectric constant
of the sample.\citep{cullen1,cullen2,TH96} The unloaded (empty)
cavity has an axisymmetric mode spacing that is $c/2L$, where $L$
is the distance
between two mirrors, also known as cavity length. In our cavity,
since $L$ is around 15 cm, the unloaded mode spacing is about 1GHz.

Putting a sample on the bottom mirror perturbs the modes in a
calculable but nontrivial way. To avoid geometrical factors, we do a
second perturbation which involves flipping the sample upside
down. Since the boundary values of the E-field are different, we are
able to get a simple (geometry-free) formula for the complex
permittivity from 3 sweeps around a 00q mode. This complex
dielectric permittivity is related to total electrical/optical
conductivity and absorption coefficient\citep{ST77} with the use of
basic theory of electromagnetism as
\begin{equation*}
n_{re} \alpha = 120 \pi \sigma_{re} = 30 \omega \epsilon_{im}
\end{equation*}
where n$_{re}$ and $\sigma_{re}$ are the real part of the refractive
index and conductivity (in $\Omega^{-1}$cm$^{-1}$), $\epsilon_{im}$ is
imaginary part of the complex dielectric constant, and $\omega$ and
$\alpha$ are frequency and absorption coefficient expressed in terms
wave numbers (cm$^{-1}$ relative to the speed of light), so 
1cm$^{-1}$ corresponds to 30 GHz. The electrical
conductivity is a macroscopic quantity which can be due to bound electrons (which is responsible for
polarization and relaxation) or free electrons (which is Drude-type mechanism). This is also valid for
absorption coefficient. These parameters, $\sigma_{re}$ and $\alpha$,
essentially describe the loss mechanisms in a material.

\begin{figure}
  \includegraphics[width=70mm]{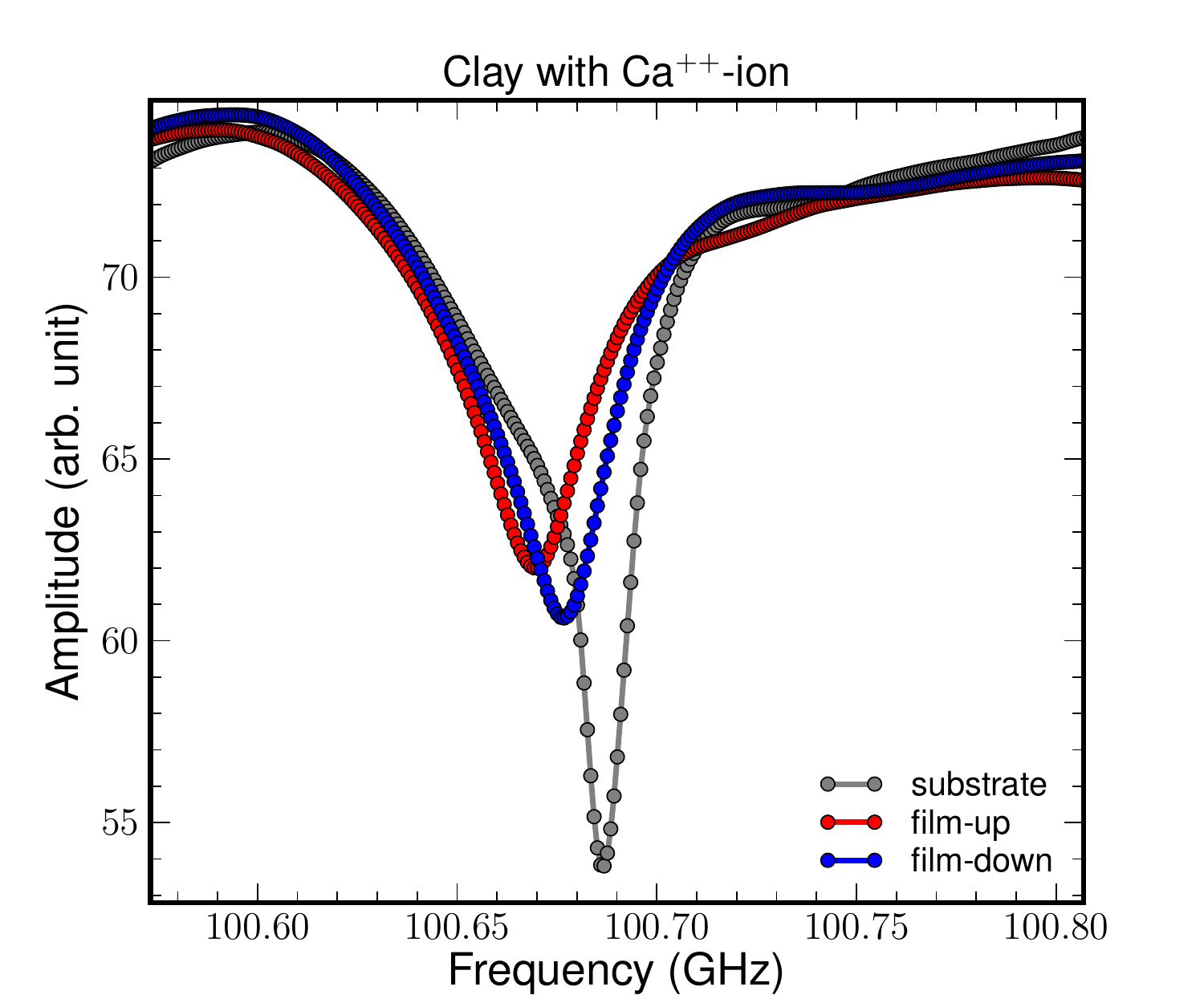}
  \label{fig:subf1}
  \includegraphics[width=70mm]{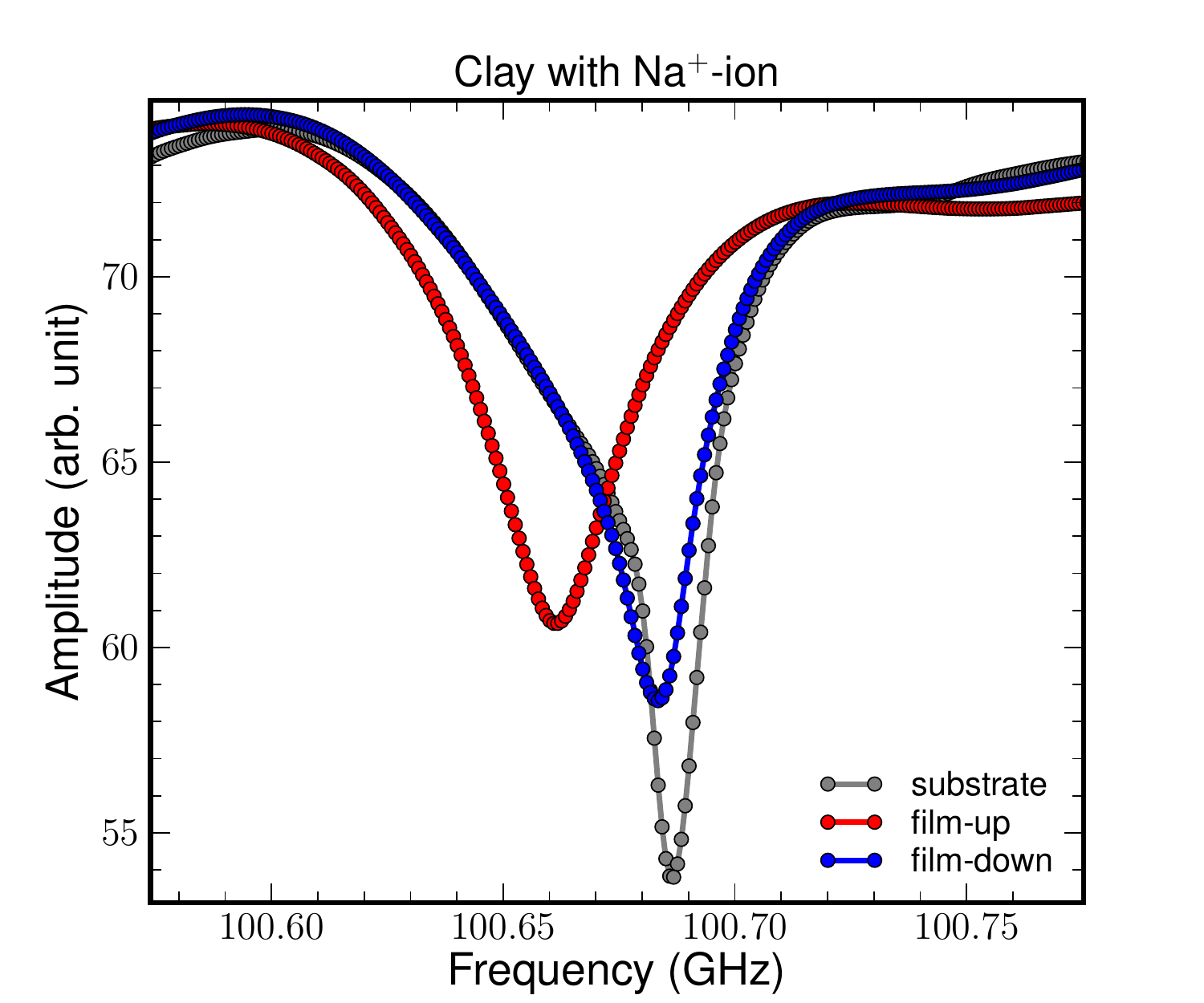}
   \label{fig:subf2}
   \caption{Perturbations with substrate only, film up, and film down set-ups
             for clay with (a) Ca$^{++}$ and (b)Na$^{+}$-ions (at room temperature)}.
\label{fig:f12}
\end{figure}

\subsection{Measurements}
   By sweeping the VNA, we identified the axissymetric empty cavity
   modes based on constant frequency-spacing.  This fixes the cavity-length,
   L, to be 145.56 mm and kept it unchanged throughout the experiments.
   Using a least-squares fit of the data to a Breit-Wigner model (as described in \citet{RR13a}) we
   retrieve the eigenfrequency-shifts and modal quality factors
   (Q-values related to a linewidth) for substrate-only, film up and
   film down positions in order to apply the differential method. In
   order to determine uncertainty in the experiment, we repeat the
   entire procedure of inserting the sample, performing the
   measurements, and taking it out six times and calculating the
   variations in frequency shifts and linewidth changes of
   substrate. In \citet{RR13a} we showed that by re-doing
   the entire procedure six times for borosillicate glass
   substrate, we obtained standard deviations less than 1.0\%
   in the complex dielectric constant.  This uncertainty also
   propagates to the calculations of dielectric constant.  For
   consistency we confirmed during each trial that the same part of the samples is probed.

The frequency shifts due to film up, film down, and substrate only 
are used in the Eqs.(\ref{equ:differentialtech}) to
determine the real part of the refractive index of a thin film.\citep{RR13a,Dudorov95}
This is called the differential method since it involves measurements with the sample film
up and down; the film will then see a different field because of the boundary 
conditions.  The result is that sample geometry factors cancel out, greatly
simplifying the calculation. Let $n_{f}$ and $n_{s}$ refer, respectively to the 
refractive index of the film and substrate being measured. Then one 
can show that
\begin{equation}
\frac{\delta\nu_{f}}{\delta\nu_{s}} =  \frac{n_{f}^{2} - 1}{n_{s}^{2} - 1}.\label{equ:differentialtech}
\end{equation}
With
\begin{equation}
 \delta\nu_{f} = \nu_{(fup)} - \nu_{(s)}.\label{fupshift}
\end{equation}
\begin{equation}
 \delta\nu_{s} = \nu_{(fdown)} - \nu_{(s)}.\label{fdownshift}
\end{equation}
Here, $\nu_{(fup)}$ , $\nu_{(fdown)}$ , $\nu_{(s)}$ represent the
eigenfrequency associated with the film on the top (film up), film at
the bottom (film down), and the substrate only, respectively.  The
term, $\delta\nu_{s}$, stands for the difference between the
eigenfrequencies associated with film at the bottom of the substrate
and the substrate only, and the term, $\delta\nu_{f}$, is the
difference between the eigenfrequencies with the film at the top of
substrate and the substrate only.  The required condition is the real
part of the refractive index of substrate must be known. (Or since
the substrate is large enough, we can measure its permittivity
directly using the quasi-optical methodology.  The real
part of the dielectric constant of thin film can obtained by
$\epsilon_{re}^{(f)}$=(n$_{re}^{(f)})^{2}$.  We need Q-values of film
up and substrate (only) to calculate the imaginary part of complex
dielectric constant of the thin film.  The Q-value of a resonant peak
(perturbation) is defined as Q=$\nu_{0}$/$\Delta\nu$. The Q-value is
related to the imaginary part of the refractive index by
n$_{im}$=1/2Q, so for a thin film, n$_{im}^{(f)}$=1/2Q$^{(f)}$.
The latter expressions assume that Q is not too small (say less than
10).  In a cavity such as ours, the Q could be $10^4 - 10^5$ depending
on technical details of the waveguide coupling.   
In the differential method, Q$^{(f)} = ( $ Q$^{(film up)} - $ Q$^{(substrate)} )$. 
Now, we are able to compute the imaginary part of the complex dielectric
constant of the thin film by 
$$\epsilon_{im}^{(f)} = 2\hbox{\rm n}_{re}^{(f)} \hbox{\rm n}_{im}^{(f)}$$.

\begin{figure}
 \centerline{
     \includegraphics[width=140mm]{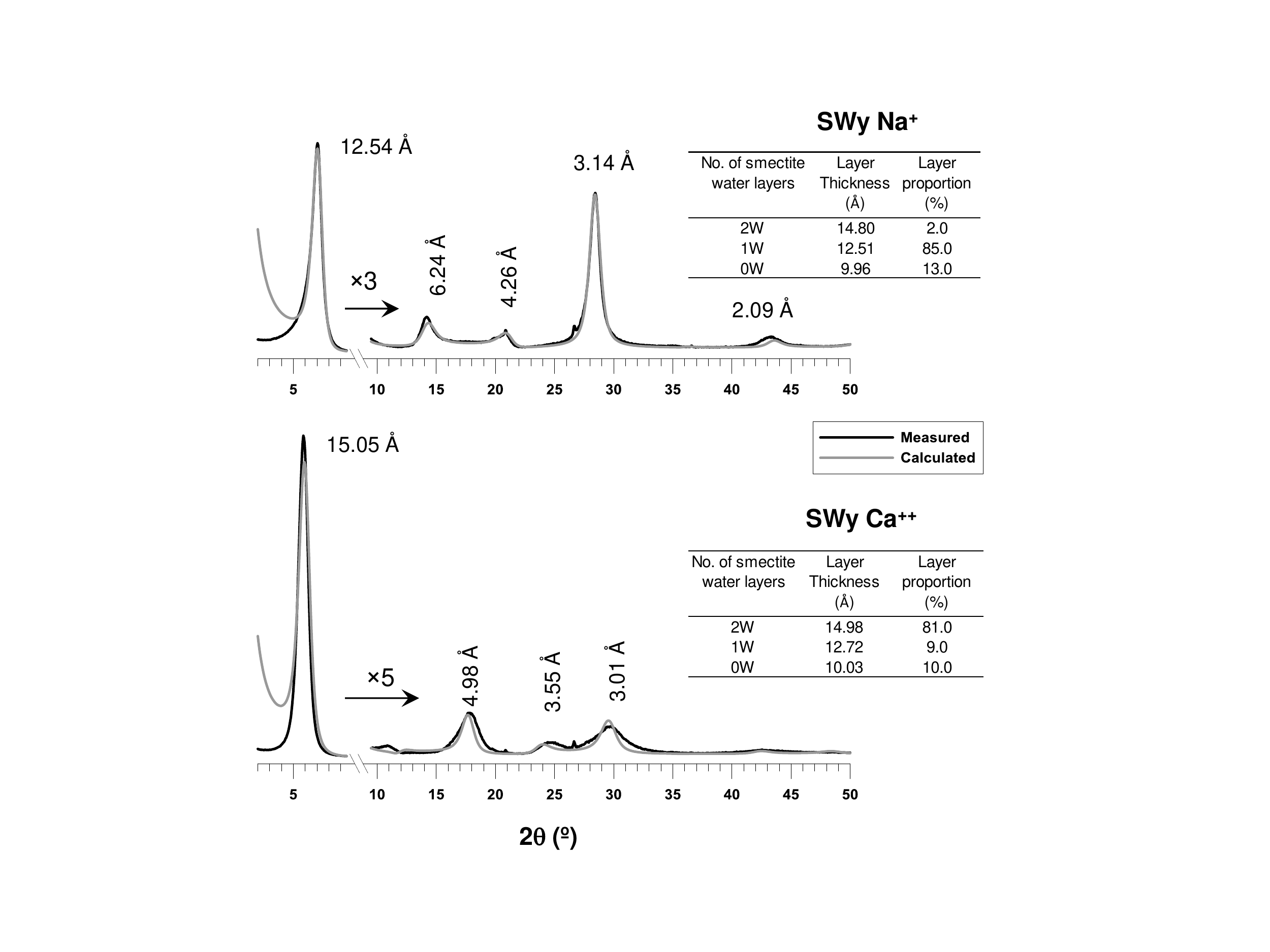}}
     \caption{Experimental and simulated XRD patterns from oriented aggregate preparations from Ca$^{++}$ and Na$^{+}$-ion stabilized
              clay samples used in EM analysis.}
\label{fig:2}
\end{figure}

\subsection{Samples and Sample preparations}
We studied smectite clay minerals (SWy) from the Clay Minerals
Society ($\|www.clays.org\|$) that were treated to yield homoionic, 
univalent (Na$^{+}$) and divalent (Ca$^{++}$) samples.  
The samples were treated to remove
carbonate and iron oxide cements with a Na-acetate buffer and
Na-dithionite respectively.\citep{JML85} The sub-0.5 $\mu$m equivalent
spherical diameter size fractions were separated from bulk samples by
standard centrifuge methods. Then the Na$^{+}$ saturated clay was
thoroughly cleaned with dialysis to remove excess salt.  To prepare
the Ca$^{++}$ exchanged smectite, a dialyzed Na$^{+}$ sample was
treated with a 1M solution of CaCl$_{2}$, and shaken for at least two
hours, excess solution was decanted and the process was repeated twice
more.  The excess Ca$^{++}$ salt was removed with
dialysis.\citep{MDM97} Oriented aggregates were made by evaporation
onto glass slides to provide a sample $\sim$ 4 cm long with at least
10 mg/cm$^{2}$ of clay.\citep{MDM97}This evaporation method yields deposition of
clay particles on the glass slide with the clay layers parallel to the glass slide.\citep{MP02}

\begin{figure}
 \centerline{
     \includegraphics[width=150mm]{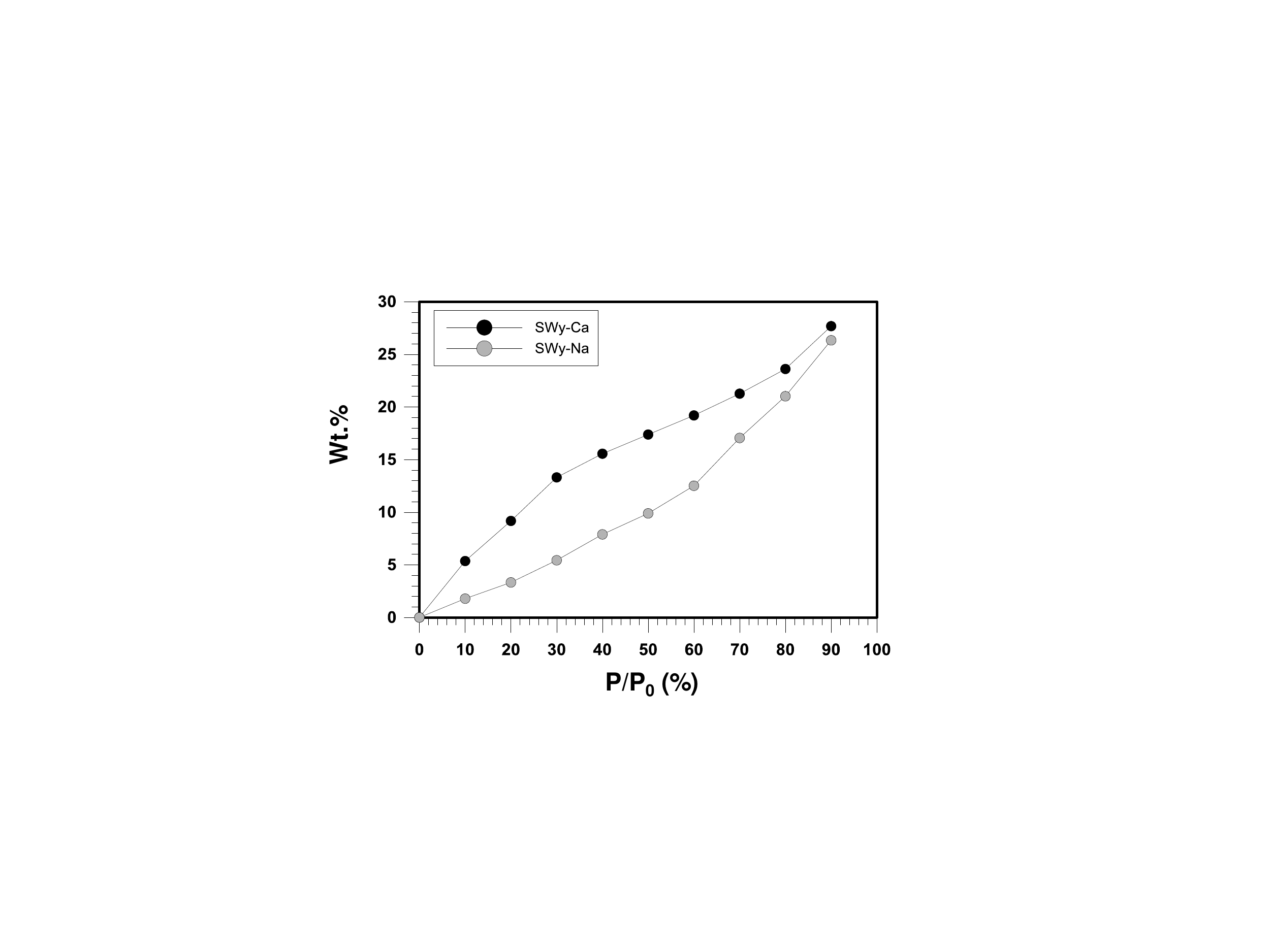}}
     \caption{Gravimetric water adsorption isotherms for SWy-Ca and SWy-Na samples following drying at 60ºC for 3 hours.}
\label{fig:3}
\end{figure}

XRD measurements showed systematic differences between Na$^{+}$ smectite and Ca$^{++}$ smectite.
The structure of smectite determined from oriented aggregate sample preparations and X-ray diffraction (XRD) 
profile modelling and especially the interlayer water complex under different environmental conditions, is described 
in detail by \citet{Ferr05a, Ferr05b, Ferr10, Ferr11}. \citet{Ferr11} discuss the influence that the structure and organization of 
interlayer water has on the material’s dielectric constant.  For this reason, the diffraction profiles from the Na$^{+}$ and Ca$^{++}$ exchanged 
preparations used in the dielectric measurements in this study were simulated following the same methods as described by \citet{drits1976} and 
\citet{drits1990}. The experimental and modelled diffraction patterns for the Na$^{+}$ and Ca$^{++}$-ion stabilized clay 
samples are shown in FIG.\ref{fig:2} along with the thicknesses of bihydrated (2W), monohydrated (1W) and dehydrated (0W) layers and their 
proportions that were used in the simulated diffraction profiles.  Under the ambient conditions of $\sim$ 50$\%$ relative himidity (RH) in the 
laboratory the Na$^{+}$ smectite is well described by random interstratifications of 2W (2$\%$), 1W (85$\%$), and 0W (13$\%$).  
In contrast, the hydrated structure of Ca$^{++}$ smectite was modelled by a different arrangement of layer thicknesses and proportions where 2W (81$\%$), 
1W (9$\%$), and 0W (10$\%$) were randomly interstratified.  The XRD simulations in FIG.\ref{fig:2} are not perfect with some misfit 
present at different angular ranges for the Ca$^{++}$ and Na$^{+}$ samples.  This suggests that there is greater heterogeneity in the layer assemblage. However, 
the near perfect fit of the 001 reflection at 12.54 \AA{} (Na) and 15.05 \AA{} (Ca), and the generally good fit at higher angles indicate that 
the models represent the primary structures, with 1W layers being dominant in the Na$^{+}$ sample and 2W layers being the dominant layer 
type in the Ca$^{++}$ sample. Smectite interlayer water as probed by XRD, considered crystalline water,\citep{Ferr11} may or may not all 
be coordinated to interlayer cations.  Based on a combination of XRD and Grand-Canonical Monte Carlo simulations\citep{Ferr11} 
found that fluctuations of charge locations and water dipoles affect the dielectric constant.  These authors also point out that gravimetric adsorption 
methods produce similar, but slightly higher molar equivalent water content than XRD experiments conducted under the same humidity conditions and based 
on the XRD simulation method.  This is due to additional adsorption on surfaces and condensation in nm-scale pores.

Thus, in addition to millimeter wave EM analysis, we performed a variety of measurements on the two samples, including thermal gravimetric analysis (TGA), 
gravimetric water adsorption, and subcritical nitrogen gas adsorption (SGA) to characterize physical differences that exist between the Na$^{+}$ and Ca$^{++}$ 
ex-infused samples. Na$^{+}$ smectite has higher SSA and dominant 1-3 nm pores typical for clay 
aggregates; in contrast, the Ca$^{++}$ smectite sample has predominantly larger pores between 50-100 nm.\citep{Kuila2014}
At $\sim$ 50 RH gravimetric water adsorption data (FIG. \ref{fig:3}) from SWy-Ca$^{++}$ and SWy-Na$^{+}$ showed that the Ca$^{++}$ smectite 
adsorbed $\sim$ 17.5 wt.$\%$ following dehydration at 60$^{0}$C for 3 hours, compared to 10 wt.$\%$ for the Na$^{+}$ smectite under the same conditions. 
The hydration enthalpy is about four times greater for Ca$^{++}$ than Na$^{+}$ cations and is consistent with these gravimetric data. 
However, in the oriented aggregate preparations used for the EM analysis the contrasting structural arrangement of the interlayer 
crystalline water between the Na$^{+}$ and Ca$^{++}$ smectite samples (FIG. \ref{fig:2}) must play a significant role in the EM response.\citep{Ferr11}

\begin{figure*}
  \includegraphics[width=60mm]{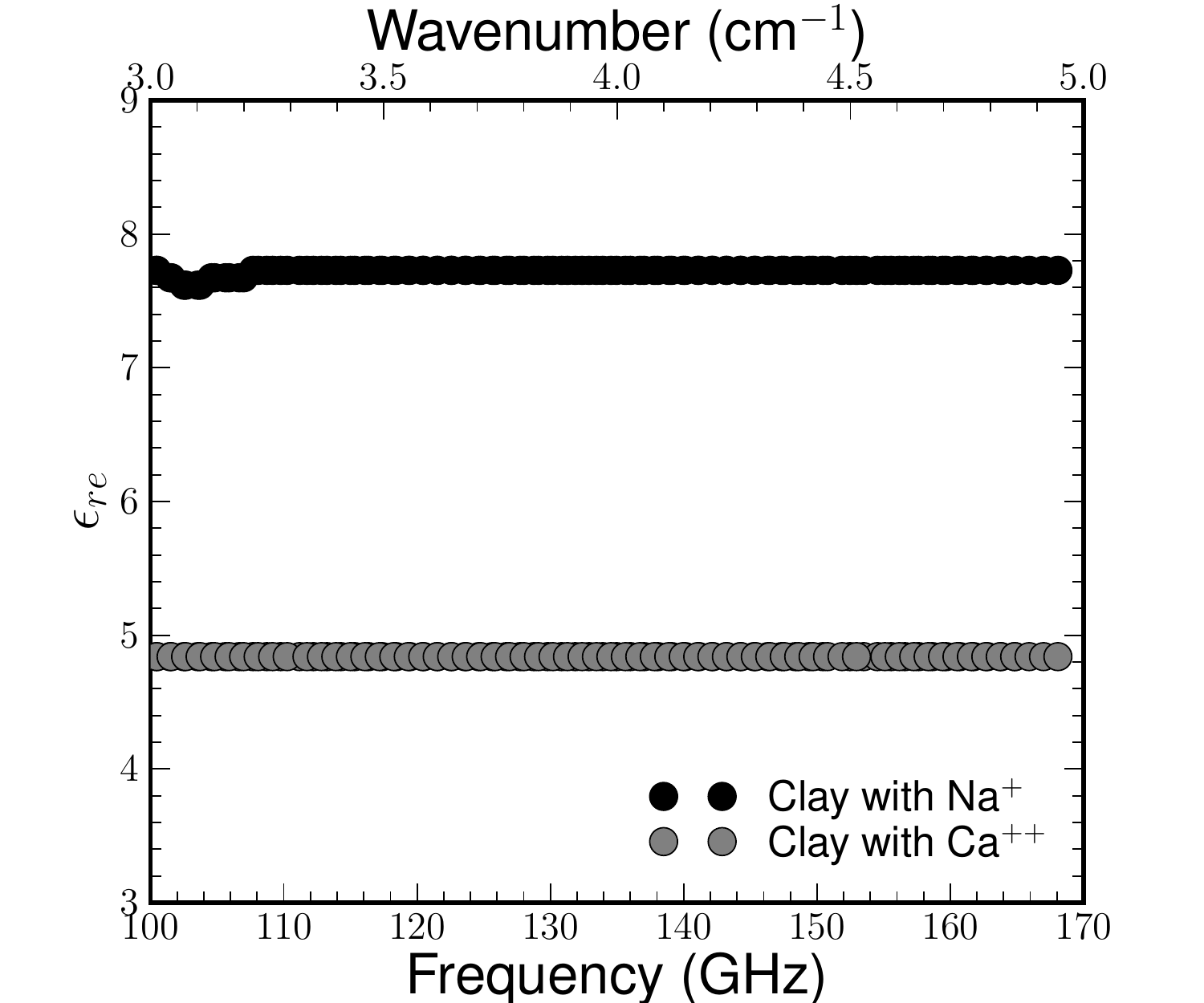}
  \label{fig:subf3}
\includegraphics[width=60mm]{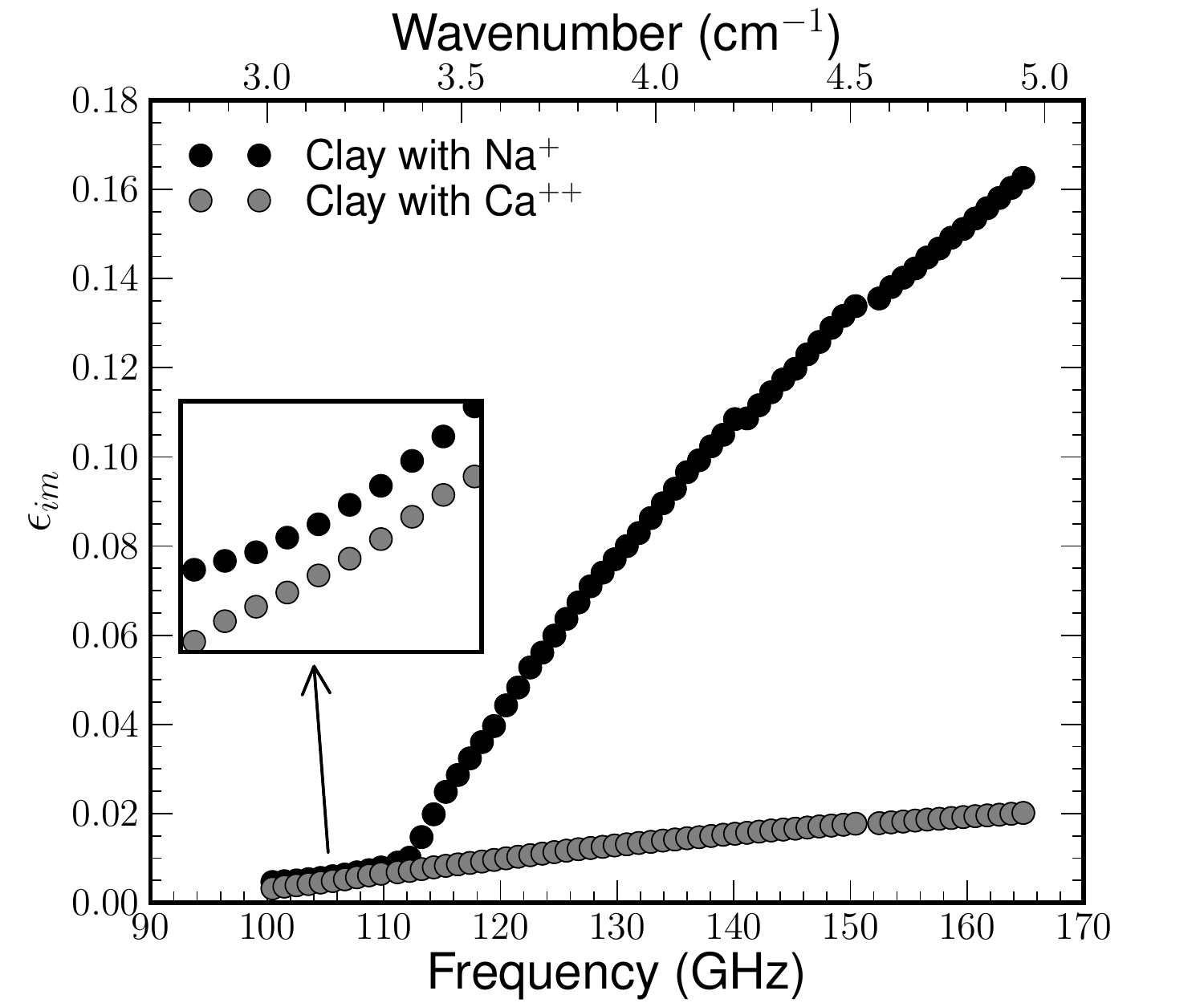}
   \label{fig:subf4}
\includegraphics[width=60mm]{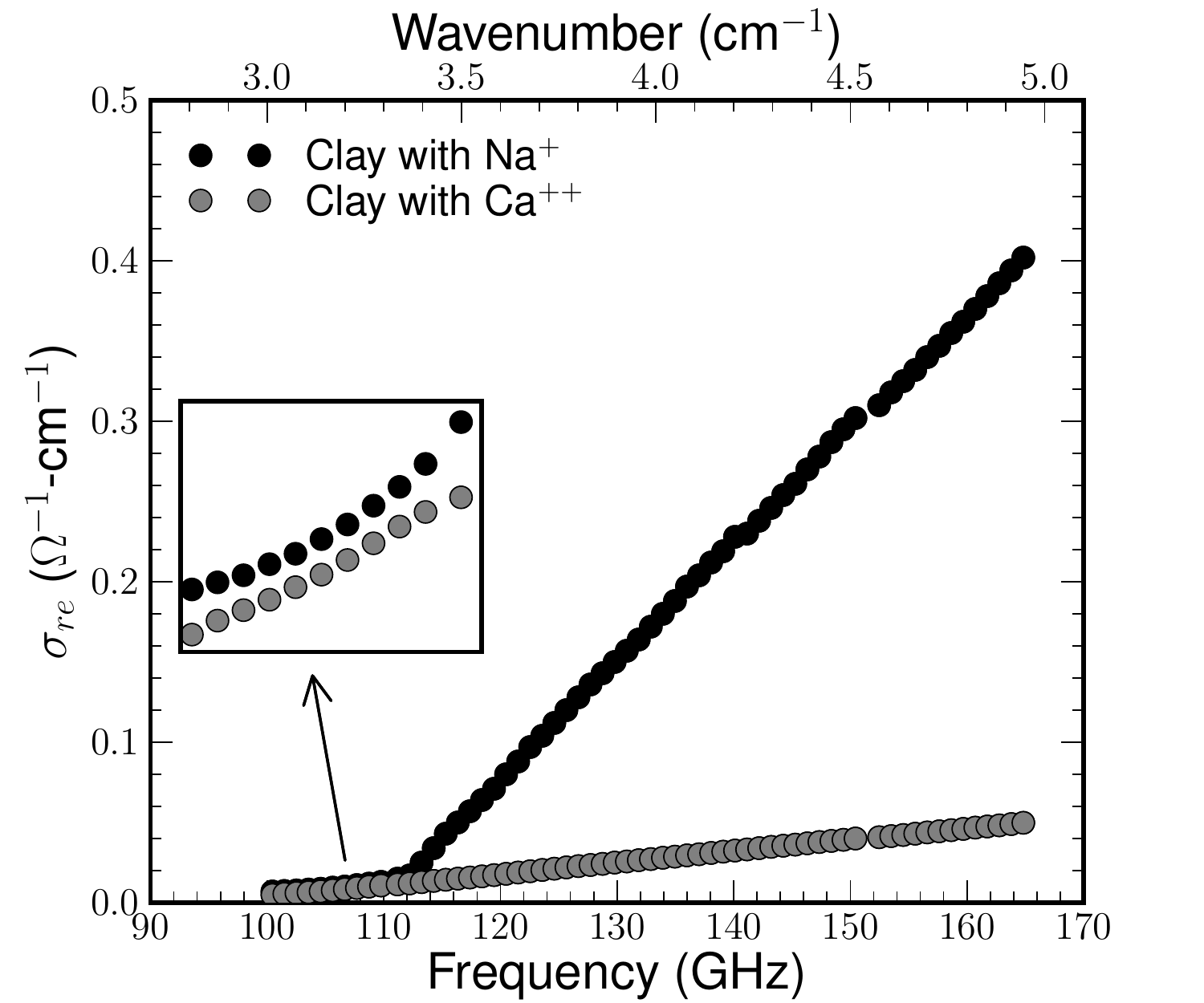}
   \label{fig:subf5}
 \caption{High frequency (100-165 GHz) data of (upper left) real part and (upper right)
   imaginary part of the complex dielectric constants of clay samples
   with Na$^{+}$/Ca$^{++}$-ions (at room temperature); (lower middle)
   conductivity of both samples (inner plot is the expansion of the
   100-110GHz responses).}
\label{fig:f345}
\end{figure*}

\section{Results and Discussion}
Referring to FIG.\ref{fig:f345}, one can
see that the real part of the dielectric constants of both clay samples
  are nearly dispersion free (i.e., frequency independent) in our
 range of measurements, indicating that carrier
  concentrations are low in both.\citep{ZLN83,RR14} The imaginary part
  of dielectric constant and electrical conductivity of Na$^{+}$-ionized sample,
  increase at two different rates: faster in the higher
  frequency range of our measurement
  and slower at low frequencies. The presence of both Debye
  relaxation\citep{PD29}, and Maxwell-Wagner relaxation\cite{BH53} at
  radio frequencies has been reported.\citep{RC75}
  But between 100 and 165 GHz, phonon
  induced relaxation is dominant.\citep{ZLN83}

 \citet{AR13} proposed a theoretical model to study complex
  conductivity-dependences on cation exchange capacity (CEC), specific
  surface area (SSA), and salinity for clay samples at low frequency.
  This model also relates CEC to SSA with
  consistency.\citep{AR13} It is reported that SSA increases
  exponentially for Na$^{+}$-clay and linearly for Ca$^{++}$-clay
  sample.\citep{KE81} From our research, it is evident that
  conductivity depends on CEC or SSA. The larger CEC or
  expnonetial-growing SSA can contribute to more disperse Stern layer
  and the smaller CEC or linearly-progressive SSA stabilize the Stern
  layer.  Thus, electrical conductivity is linked to CEC 
  and zeta potential in the Stern layer.\citep{MC95} The
  Na$^{+}$ makes a thicker unstable double layer where these high
  mobility ions are able to polarize rapidly. Therefore, the
  conductivity is more dispersive. This interlayer
  polarization is correlated to relaxation mechanisms.  The relaxation
  processes are, therefore, dependent of $\zeta$-potential which is
  also correlated to CEC.\citep{DZ10} On the other hand, The imaginary
  part of dielectric constant and conductivity of the sample with
  Ca$^{++}$ increase monotonically and sublinearly.  The Ca$^{++}$
  creates more stable double Stern layer. Due to low mobility, the
  interlayer polarization is less disperse so its conductivity is
  sublinear. Since, $\epsilon_{im}$ = 2 (1/$\nu$) $\sigma_{re}$, the
  effect of (1/$\nu$) is more into imaginary part of complex
  dielectric constant of samples with Na$^{++}$ than that of
  Ca$^{++}$. In this case, $\epsilon_{im}$ at high frequencies faced
  steeper decrease than at lower frequency ends.

Comparing the results from \citet{RT86} with ours (FIG.\ref{fig:f345}), it is clear that in the sub-THz, the
electrical conductivities of clay-Na$^{+}$/Ca$^{++}$-ions are almost 2 to 3
orders of magnitude greater than in the RF/microwave range.
On the other hand, the real part of dielectric constants for both samples,
are roughly an order of magnitude larger in the RF/microwave range than
in the sub-THz. 
The huge
RF/microwave permittivities have
been attributed to polarization of the double-layer surrounding the
ions.\citep{RT86}
The lower (and nearly frequency-independent) values of $\epsilon_{re}$
suggest that there are depletions of mobile charges due to phonon
interactions causing the diffused layers to thin.

\section{Conclusion}
  We measured sub-THz complex dielectric properties of clay samples
  with Na$^{+}$/Ca$^{++}$, and compute their $\epsilon_{re}$ and
  $\sigma_{re}$.  We illustrate the connections between
  electromagnetic parameters and components of surface chemistry such
  as CEC or SSA, $\zeta$-potential.  This also enables us to study the
  clay content and free radicals in shales, and to investigate their
  CEC and $\zeta$-potential-dependences.  In future, we will study
  more different ionized clay samples and broader frequency ranges to
  capture more complete frequency-dependences of $\epsilon_{re}$ and
  $\epsilon_{im}$. This will allow us to model computationally the
  dispersions of these parameters.

\section{Acknowledgement}

This work was supported by OCLASSH consortium and the US Department of
Energy (Basic Energy Science) under grant DE-FG02-09ER16018. We thank
Dr. John Scales for his insightful discussion on complex dielectric measurements.
The Scales lab makes all of its data available, either on-line or by
request; at no cost.  This work is dedicated to the memory of Mike
Batzle, who inspired us all.


%

\end{document}